# Thermal molecule and atom test of the modified special relativity theory

Jian-Miin Liu*
Department of Physics, Nanjing University
Nanjing, The People's Republic of China
*On leave. Present mailing address: P.O.Box 1486, Kingston, RI 02881, USA

**Abstract**
The correction to Maxwell-Boltzmann's velocity distribution law is obtained in the framework of the modified special relativity theory. The detection of velocity and velocity rate distributions for thermal molecules or atoms can serve as a test of the modified special relativity theory.

## 1. Introduction

There is a class of equivalent inertial frames of reference, any one of which moves with a non-zero constant velocity relative to any other. Each of the inertial frames of reference is supplied with motionless, rigid unit rods of equal length and motionless, synchronized clocks of equal running rate. Einstein defined the usual inertial coordinate system $\{x^r,t\}$, r=1,2,3, for each inertial frame of reference saying: "in a given inertial frame of reference the coordinates mean the results of certain measurements with rigid (motionless) rods, a clock at rest relative to the inertial frame of reference defines a local time, and the local time at all points of space, indicated by synchronized clocks and taken together, give the time of this inertial frame of reference." [1]. Besides two fundamental postulates, (i) the principle of relativity and (ii) the constancy of the light speed in all inertial frames of reference, special relativity also uses another assumption which concerns the Euclidean structure of gravity-free space and the homogeneity of gravity-free time in the usual inertial coordinate system,

$dX^2 = \delta_{rs} dx^r dx^s$, r,s=1,2,3,
$dT^2 = dt^2$,

everywhere and every time [2-5].

In our recent work [6-9], we introduced the primed inertial coordinate system $\{x'^r, t'\}$, r=1,2,3, in addition to the usual inertial coordinate system, for each inertial frame of reference. We assumed the Euclidean structures of gravity-free space and time in the primed inertial coordinate system and their non-Euclidean structures in the usual inertial coordinate system,

$dX^2 = \delta_{rs} dx'^r dx'^s = g_{rs}(y) dx^r dx^s$,   r,s=1,2,3,    (1a)
$dT^2 = dt'^2 = g(y) dt^2$,    (1b)
$g_{rs}(y) = K^2(y) \delta_{rs}$,    (1c)
$g(y) = (1-y^2/c^2)$,    (1d)
$K(y) = \dfrac{c}{2y} (1-y^2/c^2)^{1/2} \ln \dfrac{c+y}{c-y}$,    (1e)

where dX and dT are respectively the space distance and time interval between two neighboring points $(x'^1, x'^2, x'^3, t')$ and $(x'^1+dx'^1, x'^2+dx'^2, x'^3+dx'^3, t'+dt')$ in the primed inertial coordinate system or $(x^1, x^2, x^3, t)$ and $(x^1+dx^1, x^2+dx^2, x^3+dx^3, t+dt)$ in the usual inertial coordinate system, $y^s = dx^s/dt$, s=1,2,3, is the usual (Newtonian) velocity well-defined in the usual inertial coordinate system, $y = (y^s y^s)^{1/2}$. We also combined this assumption with two postulates (i) and (ii) to modify special relativity. The modified special relativity theory involves two versions of the light speed, infinite speed c' in the primed inertial coordinate system and finite speed c in the usual inertial coordinate system. It involves the c'-type Galilean transformation between any two primed inertial coordinate systems and the localized Lorentz transformation between any two usual inertial coordinate systems. The physical principle in the modified special relativity theory is: The c'-type Galilean invariance in the primed inertial coordinate system plus the transformation from the primed inertial coordinate system to the usual inertial coordinate system. The modified special relativity theory and quantum mechanics together found a convergent and invariant quantum field theory.

In this paper, we derive, from the assumption (1a)-(1e), some equations involved in the velocity space. These equations will lead to the correction to Maxwell-Boltzmann's velocity distribution law. We propose detecting the velocity and velocity rate distributions of thermal molecules or atoms as a test of the assumption (1a)-(1e) on local structures of gravity-free space and time, as well as, of the modified special relativity theory.

## 2. Velocity space in the modified special relativity theory

From the assumption (1a)-(1e) we have

$$Y^2 = \delta_{rs} y'^r y'^s, \quad r,s=1,2,3, \tag{2}$$

and

$$Y^2 = [\frac{c}{2y} \ln \frac{c+y}{c-y}]^2 \delta_{rs} y^r y^s, \quad r,s=1,2,3, \tag{3}$$

where $Y = dX/dT$ is the velocity-length, $y'^s$ is the primed velocity defined by $y'^s = dx'^s/dt'$, $s=1,2,3$, in the primed inertial coordinate system, $y' = (y'^s y'^s)^{1/2}$.

Equation (2) embodies what equation (3) does, i.e. the velocity space in the modified special relativity theory, which is defined by

$$dY^2 = \delta_{rs} dy'^r dy'^s, \quad r,s=1,2,3, \tag{4}$$

in the primed velocity-coordinates $\{y'^r\}$, $r=1,2,3$, or by

$$dY^2 = H_{rs}(y) dy^r dy^s, \quad r,s=1,2,3, \tag{5a}$$

$$H_{rs}(y) = c^2 \delta^{rs}/(c^2-y^2) + c^2 y^r y^s/(c^2-y^2)^2, \quad \text{real } y^r \text{ and } y<c, \tag{5b}$$

in the usual velocity-coordinates $\{y^r\}$, $r=1,2,3$ [10].

This statement consists of two parts. The first part, the relationship between Eqs.(2) and (4), is obvious, while its second part, the relationship between Eq.(3) and Eqs.(5a)-(5b), is not so obvious. Several authors studied the space with metric tensor $H_{rs}(y)$ [4-5,11-14]. In the velocity space described by Riemannian structure (5a)-(5b), the geodesic line between any two points $y_1^r$ and $y_2^r$, $r=1,2,3$, is linear [5]:

$$y^r = y_1^r + \alpha(y_2^r - y_1^r), \quad 0 \leq \alpha \leq 1, \quad r=1,2,3. \tag{6}$$

Actually, with the standard calculation techniques in Riemann geometry [15], we can find

$$H^{rs}(y) = (c^2-y^2)\delta^{rs}/c^2 - (c^2-y^2) y^r y^s/c^4 \tag{7a}$$

and

$$\Gamma_{jk}^i = \begin{cases} 2y^i/(c^2-y^2), & \text{if } i=j=k; \\ y^k/(c^2-y^2), & \text{if } i=j \neq k; \\ y^j/(c^2-y^2), & \text{if } i=k \neq j; \\ 0, & \text{otherwise,} \end{cases} \tag{7b}$$

where the Christoffel symbols are

$$\Gamma_{ij}^k = H^{km}(y) \Gamma_{ij,m},$$

$$\Gamma_{ij,m} = \frac{1}{2} [\partial H_{im}(y)/\partial y^j + \partial H_{jm}(y)/\partial y^i - \partial H_{ij}(y)/\partial y^m], \quad i,j,k,m=1,2,3.$$

The equation of geodesic line is therefore

$$\ddot{y}^r + [2/(c^2-y^2)] \dot{y}^r (y^s \dot{y}^s) = 0, \quad r,s=1,2,3, \tag{8}$$

where dot refers to the derivative with respect to velocity-length. Introducing new variables

$$w^r = \dot{y}^r/(c^2-y^2), \quad r=1,2,3, \tag{9}$$

we are able to rewrite Eq.(8) as

$$\dot{w}^r = 0, \quad r=1,2,3. \tag{10}$$

It is seen that

$$w^r = \text{constant}, \quad r=1,2,3, \tag{11}$$

is a solution to Eqs.(10). Furthermore, due to Eqs.(9) and (10), we have

$$w^s y^r - w^r y^s = \text{constant}, \quad r,s=1,2,3. \tag{12}$$

Eqs.(11) and (12) specify the linear relations between any two of $y^1$, $y^2$ and $y^3$. These linear relations can be represented in terms of Eqs.(6).

We evaluate the velocity-length between two points $y_1^r$ and $y_2^r$ using

$$Y(y_1^r, y_2^r) = \int_1^2 dY$$

along the geodesic line (6), where lower limit 1 and upper limit 2 refer to $y_1^r$ and $y_2^r$ respectively. At some length, we can find

$$Y(y_1^r, y_2^r) = \frac{c}{2} \ln \frac{b+a}{b-a} \,, \tag{13a}$$

$$b = c^2 - y_1^r y_2^r, \quad r=1,2,3, \tag{13b}$$

$$a = \{(c^2 - y_1^i y_1^i)(y_2^j - y_1^j)(y_2^j - y_1^j) + [y_1^k(y_2^k - y_1^k)]^2\}^{1/2}, \quad i,j,k=1,2,3. \tag{13c}$$

In the case of $y_1^r = 0$ and $y_2^r = y^r$, the velocity-length is

$$Y = \frac{c}{2} \ln \frac{c+y}{c-y} \,, \tag{14}$$

where $y = (y^r y^r)^{1/2}$, $r=1,2,3$.

Eq.(14) is nothing but Eq.(3). That judges the second part of the statement.

3. **Transformation between the primed and the usual velocity-coordimates**

Equations (2) and (3) imply

$$y'^s = [\frac{c}{2y} \ln \frac{c+y}{c-y}] y^s, \quad s=1,2,3, \tag{15a}$$

and

$$y' = \frac{c}{2} \ln \frac{c+y}{c-y} \,, \tag{15b}$$

while equations (4), (5a) and (5b) imply

$$dy'^r = A^r_s(y) dy^s, \quad r,s=1,2,3, \tag{16a}$$

$$A^r_s(y) = \gamma \delta^{rs} + \gamma(\gamma-1) y^r y^s / y^2, \tag{16b}$$

where

$$\gamma = 1/(1 - y^2/c^2)^{1/2}, \tag{17}$$

because

$$\delta_{rs} A^r_p(y) A^s_q(y) = H_{pq}(y), \quad p,q=1,2,3. \tag{18}$$

In the velocity space, the Galilean addition law among primed velocities is linked up with the Einstein addition law among usual velocities. To prove this, we let IFR1 and IFR2 be two inertial frames of reference, where IFR2 moves relative to IFR1 with usual velocity $u^r$, $r=1,2,3$, as measured in IFR1. We further let an object move relative to IFR1 and IFR2 with usual velocities $y_1^r$ and $y_2^r$, $r=1,2,3$, respectively, as measured in IFR1 and IFR2. Three primed velocities are $y_1'^r$, $y_2'^r$, $u'^r$, $r=1,2,3$, respectively corresponding to $y_1^r$, $y_2^r$, $u^r$.

The Galilean addition law among three primed velocities reads

$$y_2'^r = y_1'^r - u'^r, \quad r=1,2,3. \tag{19}$$

For IFR1, using Eqs.(2), (7a), (7b) and (7c), we have

$$[(y_1'^r - u'^r)(y_1'^r - u'^r)]^{1/2} = \frac{c}{2} \ln \frac{b+a}{b-a} \equiv \rho_1$$

with

$$b = c^2 - y_1^i u^i,$$
$$a = \{(c^2 - u^i u^i)(y_1^j - u^j)(y_1^j - u^j) + [u^k(y_1^k - u^k)]^2\}^{1/2},$$

or equivalently,

$$c^2 \tanh^2(\rho_1/c) = c^2 \{(c^2 - u^i u^i)(y_1^j - u^j)(y_1^j - u^j) + [u^k(y_1^k - u^k)]^2\}/(c^2 - u^i u^i)^2. \tag{20}$$

For IFR2, using Eqs.(2), (7a), (7b) and (7c), we have

$$(y_2'^r y_2'^r)^{1/2} = \frac{c}{2} \ln[(c+y_2)/(c-y_2)] \equiv \rho_2,$$

or equivalently,
$$y_2^2 = c^2 \tanh^2(\rho_2/c). \tag{21}$$
Eq.(19) combines with Eqs.(20) and (21) to give rise to the Einstein addition law among $y_1^r$, $y_2^r$ and $u^r$,
$$y_2^r = \sqrt{1-\frac{u^2}{c^2}}\left\{(y_1^r - u^r) + \left(\frac{1}{\sqrt{1-\frac{u^2}{c^2}}} - 1\right)u^r \frac{u^s(y_1^s - u^s)}{u^2}\right\}/\left[1 - \frac{y_1^k u^k}{c^2}\right], \quad r,s,k=1,2,3. \tag{22}$$

4. **The correction to Maxwell-Boltzmann's velocity distribution law**

The Euclidean structure (4) of the velocity space in the primed velocity-coordinates $\{y'^r\}$, r=1,2,3, convinces us that Maxwell-Boltzmann's velocity and velocity rate distribution formulas are valid in the primed velocity-coordinates, namely
$$P(y'^1, y'^2, y'^3)dy'^1 dy'^2 dy'^3 = N\left(\frac{m}{2\pi K_B T}\right)^{3/2} \exp\left[-\frac{m}{2K_B T}(y')^2\right] dy'^1 dy'^2 dy'^3 \tag{23}$$
and
$$P(y')dy' = 4\pi N\left(\frac{m}{2\pi K_B T}\right)^{3/2} (y')^2 \exp\left[-\frac{m}{2K_B T}(y')^2\right] dy', \tag{24}$$
where N is the number of molecules or atoms, m their rest mass, T the temperature, and $K_B$ the Boltzmann constant. We can employ Eqs.(15a), (15b), (16a) and (16b) to represent these two formulas in the usual velocity-coordinates $\{y^r\}$, r=1,2,3.

Using Eq.(15b) and
$$dy'^1 dy'^2 dy'^3 = \gamma^4 dy^1 dy^2 dy^3 \tag{25}$$
which is deduced from Eqs.(16a) and (16b), where $\gamma$ is in Eq.(17), we have from Eq.(23)
$$P(y^1, y^2, y^3) dy^1 dy^2 dy^3 = N \frac{(m/2\pi K_B T)^{3/2}}{(1-y^2/c^2)^2} \exp\left[-\frac{mc^2}{8K_B T}\left(\ln\frac{c+y}{c-y}\right)^2\right] dy^1 dy^2 dy^3. \tag{26}$$
Using Eq.(15b) and
$$dy' = \gamma^2 dy \tag{27}$$
which comes from differentiating Eq.(15b), we have from Eq.(24)
$$P(y)dy = \pi c^2 N \frac{(m/2\pi K_B T)^{3/2}}{(1-y^2/c^2)} \left(\ln\frac{c+y}{c-y}\right)^2 \exp\left[-\frac{mc^2}{8K_B T}\left(\ln\frac{c+y}{c-y}\right)^2\right] dy. \tag{28}$$
It is a condition for getting Eqs.(26) and (28) that temperature T is invariant under the transformation from the primed inertial coordinate system to the usual inertial coordinate system. The reasons for this will be given in another place [16].

Eqs.(26) and (28) are the correction to Maxwell-Boltzmann's velocity and velocity rate distributions in the usual velocity-coordinates $\{y^r\}$, r=1,2,3. It is a characteristic that both $P(y^1,y^2,y^3)$ and $P(y)$ go to zero when y approaches c. It is also a characteristic that two distribution functions $P(y^1,y^2,y^3)$ and $P(y)$ respectively reduce to their previous forms,
$$N\left(\frac{m}{2\pi K_B T}\right)^{3/2} \exp\left[-\frac{m}{2K_B T}(y)^2\right] \tag{29}$$
and
$$4\pi N\left(\frac{m}{2\pi K_B T}\right)^{3/2} (y)^2 \exp\left[-\frac{m}{2K_B T}(y)^2\right], \tag{30}$$
only when velocity is very small.

We propose to detect the velocity and velocity rate distributions of thermal molecules or atoms for testing the assumption (1a)-(1e) and the modified special relativity theory.

**Acknowledgement**

The author greatly appreciates the teachings of Prof. Wo-Te Shen. The author thanks Prof. M. S. El Naschie for his support of this work.